\documentclass[aip,reprint,pop]{revtex4-1}

\usepackage{graphicx}
\usepackage{subcaption}
\usepackage{amsmath}

\draft

\begin{document}

\title{Characterization of the equilibrium configuration for modulated beams in a plasma wakefield accelerator} 

\author{Roberto Martorelli}
\affiliation{Heinrich Heine Universit\"{a}t, 40225 D\"{u}sseldorf, Germany}

\author{Alexander Pukhov}
\affiliation{Heinrich Heine Universit\"{a}t, 40225 D\"{u}sseldorf, Germany}

\date{\today}

\begin{abstract}
We analyze the equilibrium configuration for a modulated beam with sharp boundaries exposed to the fields self-generated by the interaction with a plasma. Through a
semi-analytical approach we show the presence of multiple equilibrium configurations and we determine the one more suitable for wakefield excitation. Once pointed out
the absence of confinement for the front of the beam and the consequently divergence driven by the emittance, we study the evolution of the equilibrium configuration
while propagating in the plasma, discarding all the others time-dependencies. We show the onset of a rigid backward drift of the equilibrium configuration and we
provide an explanation in the increasing length of the first bunch.
\end{abstract}


\maketitle

\section{Introduction}
\label{Sec.1}
Continuous efforts are lavished in the generation of high-energy electron bunches due to the wide range of applications in which are involved, from medicine to new
physics research. Together with an improvement of the current available accelerating techniques, considerable attempts have been made also in the development of new
methods to generate accelerated bunches. Among these, plasma wakefield acceleration, both laser driven\cite{Mangles, Faure, Geddes, Leemans} and particle driven
\cite{Hogan, Ian}, has received considerable attention in recent years due to the high electric fields that the plasma can sustain. A promising aspect of the particle
driven technique is the possibility to use the currently available hadron bunches produced in synchrotons, whose energies are much higher than those achieved by
electron/positron accelerators. The AWAKE project\cite{Awake} at CERN aims exactly to proof this scenario, employing the $400$ GeV/c proton bunches produced at the
Super Proton Synchrotron(SPS) as a driver for the first proton driven plasma wakefield accelerator.\par
The optimal generation of a wake requires a driver whose length is approximately $\lambda_p/2$ where $\lambda_p=2\pi c/\omega_p$ is the plasma wavelength and
$\omega_p=(4\pi ne^2/m)^\frac{1}{2}$ is the plasma frequency. Proton bunches are usually several times longer than $\lambda_p$ and such a requirement is achieved by
using the self-modulation instability\cite{Kumar}, in which a long driver interacts with its own wake. At its final stage the modulation provides a train of bunches
with the requested length whose periodicity is approximately $\lambda_p$.\par
Although the self-modulation instability provides a configuration suitable for the excitation of a strong wakefield, the stability of the modulated structure is on the
other hand not guarantee. The same transverse fields that modulate the initial bunch are responsible for its deterioration during the propagation in the plasma channel,
undermining therefore the resulting accelerating field.\par
Together with extended efforts to characterize the evolution of the modulated structure, such a problem has encouraged the development of other experimental design as
well, including a configuration involving two plasma cells\cite{Path}. In the first one the proton bunch got modulated through the self-modulation instability, while in
the second the modulated structure excites the field suitable for the acceleration of the electrons.\par
An analogous framework involves instead the injection of a pre-modulated beam directly in the plasma cell, avoiding the onset of the self-modulation instability.\par
In this paper we study the equilibrium configuration obtained by a modulated beam self-interacting with its own wakefield. The modulated beam can be obtained either by
pre-modulation or by the self-modulation instability at its final stage. The analysis is performed in the quasi-static approximation assuming a linear plasma response
generated by a hard-cut bunch.\par
The paper is structured as follows: in section II we trace the model used for the description of the beam-plasma interaction, outlining the physical effects occurring;
in section III we describe the possible equilibrium configurations for the bunch emerging from the model, underlining the differences and therefore providing the more
suitable equilibrium structure; in section IV we analyze the evolution of the chosen equilibrium configuration while propagating in the plasma channel; in section V the
conclusions are presented.

\section{Analytical model}
\label{Sec.2}
The description of the fields generated by the beam-plasma interaction is based on the results obtained in the Kenigs and Jones work\cite{Kenigs}.\par
The analysis is developed for an axi-symmetric bunch linearly interacting with a plasma with immobile ions. The plasma is overdense, therefore the beam can be treated
as an external perturbation. The system is studied in the co-moving frame defined by the variables $\xi=\beta ct-z$ and $\tau=t$ with $\beta=v_b/c\simeq1$ and $c$ the
speed of light. A further assumption is the quasi-static approximation for the bunch, providing $\partial_\tau\simeq0$.\par
The two-dimensional transverse wakefield generated inside the driver is then:
\begin{align}
 W(r,\xi)&=(E_r-\beta B_\theta)=4\pi k_p\int_0^\infty\int_0^\xi\frac{\partial\rho(r',\xi')}{\partial r'}\nonumber\\
 &r'\operatorname{I_1}(k_pr_<)\operatorname{K_1}(k_pr_>)\sin[k_p(\xi-\xi')]d\xi'dr'\label{eq1}
\end{align}
where $k_p=\omega_p/c$ is the plasma wavenumber, $\rho(r,\xi)$ is the bunch charge density, $\operatorname{I_1}$ and $\operatorname{K_1}$ are the modified Bessel
functions and $r_{</>}=\operatorname{min}/\operatorname{max}(r,r')$.\par
The analysis is developed assuming a transverse flat-top density profile for the bunch:
\begin{equation}
 \rho(r,\xi)=n_bq_b\left(\frac{r_0}{r_b(\xi)}\right)^2\operatorname{H}(r_b(\xi)-r)f(\xi)\label{eq2}
\end{equation}
where $n_b$ is the peak bunch density, $q_b$ is the bunch charge, $r_0$ is the initial bunch radius, $r_b(\xi)$ is the radius of the beam-envelope, $\operatorname{H}$ is
the Heaviside function and $f(\xi)$ is the longitudinal bunch profile. The wakefield generated by such a distribution can be therefore formulated as:
\begin{align}
 &W(r,\xi)=-4\pi k_pn_bq_br_0^2\nonumber\\
 &\begin{cases}
 \operatorname{I_1}(k_pr)\int_0^\xi f(\xi^\prime)\frac{\operatorname{K_1}(k_pr_b(\xi^\prime))}{r_b(\xi^\prime)}\sin[k_p(\xi-\xi^\prime)]d\xi^\prime\text{ for }r<r_b\\
 \operatorname{K_1}(k_pr)\int_0^\xi f(\xi^\prime)\frac{\operatorname{I_1}(k_pr_b(\xi^\prime))}{r_b(\xi^\prime)}\sin[k_p(\xi-\xi^\prime)]d\xi^\prime\text{ for }r>r_b.\label{eq3}
 \end{cases}
\end{align}
On the other hand the model does not self-consistently include the bunch dynamics. Since we are just interested in the transverse motion of the beam particles, we
couple the field equations with the beam-envelope equation for the beam radius, assuming that the transverse motion of the beam can be described by just the motion of
its boundary (water-bag model). This is a valid approximation as long as the force acting inside the beam is approximately linear like in this analysis.\par
Denoting with $r_b=r_b(\xi,\tau)$ and with $r_b'=r_b(\xi',\tau)$, the resulting equation is:
\begin{align}
 &\frac{\partial^2r_b}{\partial\tau^2}=F(\xi,\tau)=\frac{\epsilon^2c^2}{\gamma^2r_b^3}-\frac{4\pi k_pn_bq_b^2r_0^2}{m_b\gamma}\nonumber\\
 &\begin{cases}
 \operatorname{I_1}(k_pr_b)\int_0^\xi f(\xi^\prime)\frac{\operatorname{K_1}(k_pr_b^\prime)}{r_b^\prime}\sin[k_p(\xi-\xi^\prime)]d\xi^\prime\text{ for } r_b<r_b'\\
 \operatorname{K_1}(k_pr_b)\int_0^\xi f(\xi^\prime)\frac{\operatorname{I_1}(k_pr_b^\prime)}{r_b^\prime}\sin[k_p(\xi-\xi^\prime)]d\xi^\prime\text{ for } r_b>r_b'
 \end{cases}\label{eq4}
\end{align}
with $\epsilon$ being the normalized beam emittance, $\gamma=1/\sqrt{1-\beta^2}$ the beam relativistic Lorentz factor and $m_b$ the mass of the beam particles.\par
The dynamics described by the model can be readily interpreted as a charge neutralization of the bunch particles by the oscillating plasma electrons.\par
The charge neutralization weakens the Coulomb force arising from the particles in the bunch and therefore the pinching force generated by the bunch current is not
anymore balanced, leading to a periodical focusing of the bunch. The focusing force increases the bunch density providing therefore a stronger wakefield.\par
It is worth noting that the front of the bunch does not experience the effect of the wakefield, so that its evolution is the same as of a bunch expanding in vacuum
under the action of the emittance:
\begin{equation}
 r_b(\xi=0,\tau)=r_{b0}(\tau)=r_0\sqrt{1+\frac{\epsilon^2c^2\tau^2}{r_0^4\gamma^2}}.\label{eq5}
\end{equation}
This characteristic is of key importance for the future development of the analysis.\par

\section{Equilibrium configurations}
\label{Sec.3}
On the ground of the model presented in the previous section we analyze now the transverse equilibrium of a bunch interacting with a plasma. For sake of simplicity we
perform the investigation for the case of a bunch with flat-top longitudinal density profile
$f(\xi)=\left(\operatorname{H}\left(\xi-\xi_0+\frac{L}{2}\right)-\operatorname{H}\left(\xi-\xi_0-\frac{L}{2}\right)\right)$ with $\xi_0$ the center of the bunch and $L$
its length. The parameters for both the bunch and the plasma correspond to the baseline of the AWAKE experiment (so the bunch particles are protons) with the exception
of the bunch length (Table \ref{tab1}).\par
\begin{table}
 \caption{Simulation parameters for plasma and bunch.\label{tab1}}
 \begin{tabular}{| c | r |}
 \hline
 Parameter & Value\\ \hline
 Plasma density $(n_p)$ & $7\times10^{14}$ cm$^{-3}$\\ \hline
 Bunch length $(L)$ & $1.2$ cm\\ \hline
 Initial bunch radius $(r_0)$ & $0.02$ cm\\ \hline
 Peak bunch density $(n_b)$ & $4\times10^{12}$ cm$^{-3}$\\ \hline
 Bunch relativistic Lorentz factor $(\gamma)$ & 400\\ \hline
 Normalized bunch emittance $(\epsilon)$ & $3.5$ mm mrad\\ \hline
 \end{tabular}
\end{table}
An equilibrium configuration for the bunch radius is reached  when the force generated by the emittance is able to balance the focusing force exerted by the wakefield.\par
It is worth noting that for a given initial condition for the bunch, the wakefield can be too weak for triggering the bunch focusing immediately after the front of the
bunch (that as we mentioned previously does not experience any focusing field). On the other hand moving further behind the front of the bunch more and more particles
contribute to the generation of the wake that therefore becomes more intense and able to focus the bunch. It is possible to evaluate the position $\xi_p$ at which the
pinching of the bunch first occurs for a given initial conditions by setting to zero Eq.(\ref{eq4}) with $r_b(\xi,\tau)=r_{b0}(\tau)$, provided by the fact that before
the onset of the focusing force, the bunch radius has to be constant in $\xi$ and equals to that of the front. This leads to:
\begin{align}
&\frac{\epsilon^2c^2}{\gamma^2r_{b0}^3(\tau)}=\frac{4\pi k_pn_bq_b^2r_0^2}{m_b\gamma}\nonumber\\
&\operatorname{I_1}(k_pr_{b0}(\tau))\int_0^{\xi_p}\frac{\operatorname{K_1}(k_pr_{b0}(\tau))}{r_{b0}(\tau)}\sin[k_p(\xi_p-\xi')]d\xi'\label{eq6}
\end{align}
and therefore (withouth explictly expressing the dependences by $\tau$)
\begin{equation}
 \xi_p(\tau)=\frac{1}{k_p}\arccos\left[1-\frac{\epsilon^2c^2}{\omega_b^2r_0^2\gamma^2r_{b0}^2\operatorname{I_1}(k_pr_{b0})\operatorname{K_1}(k_pr_{b0})}\right]\label{eq7}
\end{equation}
with $\omega_b^2=4\pi n_bq_b^2/m_b\gamma$. The bunch length required for exciting a strong enough wakefield decreases as expected increasing the density of the bunch and as well increasing the radius at the
front of the bunch. Since the front of the bunch is always expanding while propagating in the plasma channel, according to Eq.(\ref{eq5}), the non-neutralized section
of the bunch is a decreasing function of the propagating distance in plasma.\par
After reached the position $\xi_p$ the neutralization of the charge provides a focusing force acting on the bunch. This means that the maximum radius achievable
while propagating in the plasma is the one corresponding to the free expansion in vacuum in Eq.(\ref{eq5}) for the same propagation distance. This reflects the case in
which the focusing field is zero.\par
\begin{figure}
\includegraphics[scale=.8]{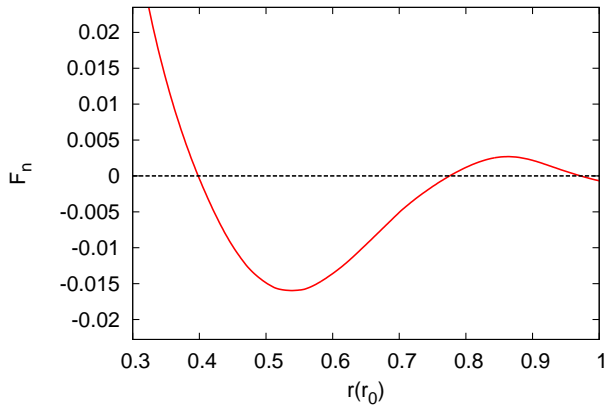}%
\caption{Normalized force $F_n=F/m_bE_0$ at $\xi=5.3\lambda_p$. $E_0=m_ec\omega_p/e$ is the wave-breaking field.\label{fig1}}%
\end{figure}

Most important, the piece-wise nature of the wakefield excited by a bunch with sharp boundaries, inevitably provides the presence of multiple roots for the equilibrium
equation, i.e. multiple equilibrium radii for the bunch at the same position $\xi$ (Fig.\ref{fig1}). This last issue implies a non unique solution for the equilibrium
radius and requires setting some prescriptions in order to choose a particular one.\par
A first assumption, in view of the focusing nature of the field, is that the equilibrium radius must be equal or smaller than the radius of the free expanding bunch
$r_b(\xi)\leq r_b(\xi=0)$. Any solution of the equilibrium radius occurring over this limit has to be discarded.\par
The second assumption is to consider just stable equilibrium configuration under small perturbations. This is a reasonable request in view of the aim to obtain a
long-lasting bunch configuration for the whole propagation distance in the plasma.\par
These two assumptions alone are not sufficient to define an equilibrium configuration for the whole bunch and an additional one has to be set according to the physical
properties we demand the bunch to posses. A possible choice consists of an equilibrium solution lying in the global minimum of the potential for every position $\xi$ in
the bunch, or else in the smaller equilibrium radius achievable.\par
In the first case we favor the stability of the equilibrium configuration, in the second one the intensity of the produced wakefield (a strongly focused driver reach
a higher density and therefore excites a stronger wave).\par
\begin{figure}
\includegraphics[scale=.8]{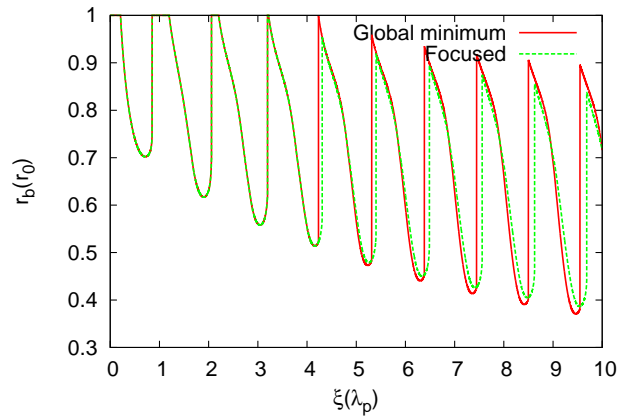}%
\caption{Normalized equilibrium beam radius for the ``global minimum case (continous line) and ``focused'' case (dashed line).\label{fig2}}%
\end{figure}
\begin{figure}
\includegraphics[scale=.8]{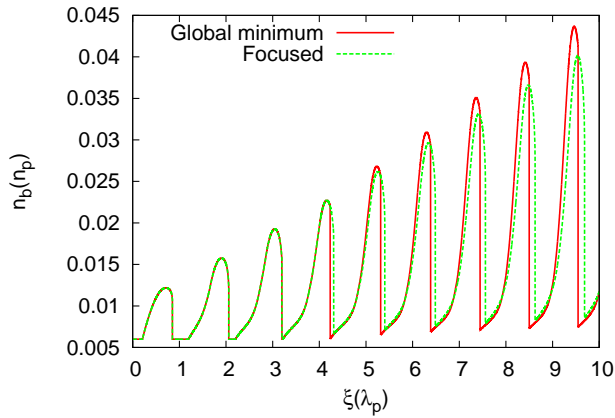}%
\caption{Equilibrium configuration for a modulated beam in the ``global minimum'' case (continous line) and ``focused'' case (dashed line).\label{fig3}}%
\end{figure}

We carry the analysis of the equilibrium structure under the previous assumptions for both the ``focused'' and ``global minimum'' cases, comparing the differences
between the two.
As shown in Fig.\ref{fig2} the bunch experiences a periodical focusing force at which corresponds consequently a modulated periodic equilibrium structure analogous
to that achieved by the onset of the self-modulation instability (Fig.\ref{fig3}). The final configuration is nevertheless different for the two choices of the
equilibrium radius. The difference emerges from the fourth bunch and represent the appearance of the third root inside the range $r_b(\xi)\leq r_b(\xi=0)$ defined
above. This leads to a shift between the two configurations that persists for the rest of the bunch.\par
The different structure for the two cases corresponds to a different behavior of the potential for every slice $\xi$ in the bunch.\par
For the case of ``global minimum'' solution the modulation occurs at the formation of a new global minimum eventually far from the previous one. Instead in the case in
which the focusing of the bunch is favored, the modulation is related to the relaxation of the potential well corresponding to the minimum radius and consequently
disappearance of one of the solution.\par
In both the configurations the peak density of the bunches grows moving towards the end of the bunch as a consequence of the constructive interference acting on the
transverse field generated by the single bunches.\par
Nevertheless we can see that the configuration corresponding to the highly focused case does not provide a higher peak density for the bunches. This is a result of the
phase shift in the transverse field respect to the ``global minimum'' case, due to the different lengths of the fourth bunch for the two configurations. The total
transverse field at every point is a result of the superposition of the fields generated by the single bunches. The different length of the forth bunch in the
``focusing'' configuration provides an imperfect matching of the phase of the transverse field, resulting in a weaker total focusing force and therefore in a lower
peak density for the bunches.\par
According to this last result therefore we can carry the rest of the work just considering the ``global minimum'' equilibrium configuration since it provides a stronger
focusing and consequently a higher peak density for the bunches as well as a more stable configuration.

\section{Evolution of the equilibrium configuration}
\label{Sec.4}
As previously mentioned, the front of the bunch is never able to reach an equilibrium position since its dynamics is controlled by its own space-charge force and not also
by the interaction with the plasma. This means that is not possible to achieve a global equilibrium for the bunch that instead experiences a slow evolution.\par
Moreover, since the dynamics at every point in the bunch depends on the upstream section, a change in the radius at the front leads inevitably to a new equilibrium
configuration also for the rest of the driver. We want to see how strongly the driver is affected by the growth of the front radius, discarding all other time dependencies
but the one that appears in Eq.\ref{eq5}.\par
\begin{figure}
\includegraphics[scale=.8]{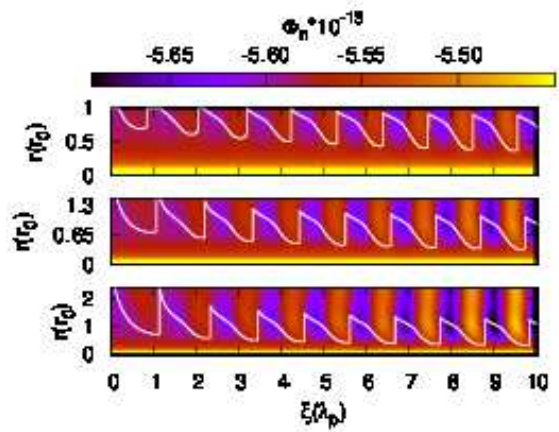}%
\caption{Potential surfaces after propagating for $0$, $5$ and $10$ meters in the plasma. The white lines corresponds to the beam radius defined by the minimum of the
potential at every position $\xi$\label{fig4}}%
\end{figure}

A qualitative behavior can be evinced by Fig.\ref{fig4} where we show the variation of the potential surfaces for a section of the bunch, due to the expansion of the
radius at the front at different propagation distances. The whole potential map appears to experience a shift toward the back of the bunch together with a global
increase in the amplitude of the potential. Such a variation implies inevitably a change in the global minima of the potential and this leads to a different
equilibrium configuration achieved while propagating in the plasma channel.\par
A similar backward drift of the potential wells has been already studied for the self-modulation instability\cite{Lotov}. In our analysis on the other hand we are able to
address as the responsible mechanism of this drift the growth of the front of the bunch due to the emittance-driven force, since all the others time dependencies of the
system have been discarded.\par
\begin{figure}
\includegraphics[scale=.8]{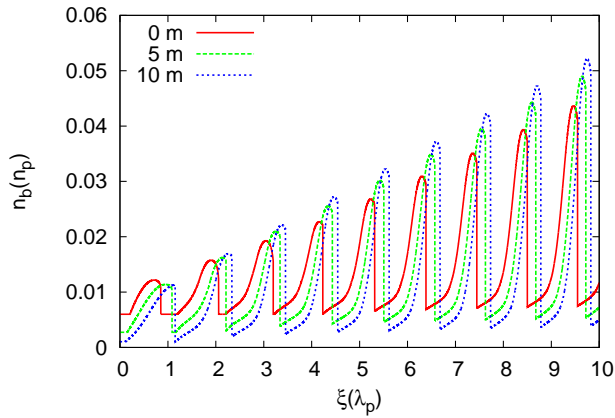}%
\caption{Equilibrium configuration after $0$ (continuos line), $5$ (dashed line) and $10$ (dotted line) meters of propagation in the plasma. The minimum densities
achieved correspond to the absence of a focusing field and therefore to a purely-expanding bunch due to the emittance-driven force\label{fig5}}%
\end{figure}

This backward shift appears more evident in Fig.\ref{fig5} where the different equilibrium configuration obtained for different propagation distances are presented. We still
obtain a modulated structure self-similar to the previous one, with periodicity and length of the bunches approximately equals the plasma wavelength, but the whole
configuration exhibits a drift towards the back of the bunch respect to those at previous propagation distances. Moreover there is as well an increase of the peak
densities of the bunches for increasing propagation distances.\par
Left out from this behavior is on the other hand the first bunch. The peak density is in fact decreasing during the propagation and also the bunch length is not
constant but it is increasing.\par
\begin{figure}
\includegraphics[scale=.8]{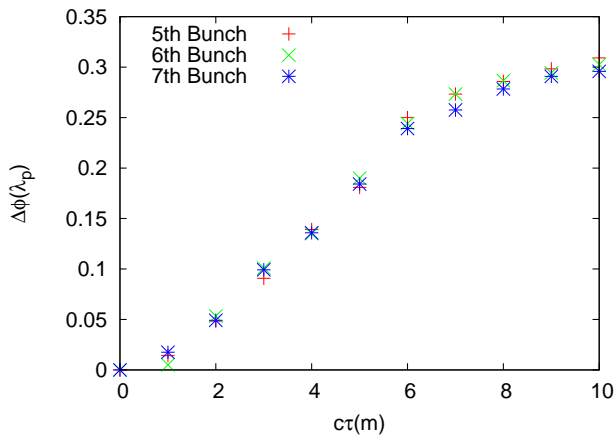}%
\caption{Shift of the $5$th, $6$th and $7$th bunch respect to its initial position for different propagation distance.\label{fig6}}%
\end{figure}

In order to characterize the evolution of the equilibrium structure and understand the reason for such a behavior we track the position of the peaks of the bunches for
different propagation distances. The first feature that emerges from Fig.\ref{fig6} is an approximately equal variation of the bunch positions respect to their initial one
for all the bunches. This means that the equilibrium configuration rigidly moves backward while propagating in the plasma and that the period of the modulation therefore
does not change. Moreover the variation is initially a linear function of the propagation distance, becoming smoother after approximately $7$ m.\par
As we pointed previously the divergence of the beam driven by the emittance is the only time-dependent mechanism in this analysis, therefore the reason for the presence of
such a drift has to be found there.\par
An interesting difference that we mentioned above in the evolving equilibrium configuration, is the anomalous behavior of the first bunch respect to the rest of
the modulated beam. In particular the length of the first bunch is increasing while propagating in the plasma.\par
\begin{figure}
\includegraphics[scale=.8]{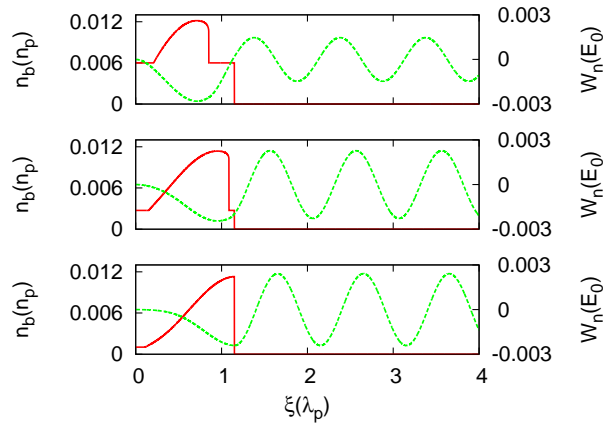}%
\caption{Bunch density and transverse wakefield at $r=r_0/2$ after $0$, $5$ and $10$ meters of propagation for a short bunch.\label{fig7}}%
\end{figure}

There are two main reasons for such a phenomena related each other. The first one is the decrease of the required bunch length for the onset of the beam focusing as
expressed previously in Eq.\ref{eq7}. The second is the longer bunch required to excite a wake as strong as for shorter propagation distances due to the decreasing bunch
density after the beam expansion. This last issue is responsible for a phase shift of the transverse field and therefore of a longer length of the first bunch (Fig.\ref{fig7}).\par
It is exactly this phase shift in the transverse field for the first bunch that causes then a backward drift of the entire equilibrium configuration. Moreover although
the maximum field inside the first bunch is a decreasing function of the propagation distance, the different length results in an increasing maximum field outside the bunch.
This results in a stronger focusing force and therefore in the increasing peak density of the bunches observed.

\section{Conclusion}
\label{Sec.5}
Summarizing, we have analyzed the transverse equilibrium configuration for a beam with sharp boundaries self-interacting with a plasma. This picture is analogous to the
self-modulation instability framework.\par
Through a semi-analytical method we have derived an expression for the required bunch length that provides the onset of the focusing force. Moreover we find the
presence of multiple equilibrium configuration achievable by the bunch due to the piece-wise nature of the wakefield generated. Analyzing the differences between the
possible modulated structure we have outlined the one that guarantees the most stable and focused bunch.\par
We pointed out how the inability to confine the front of the bunch, and the self-interacting nature of the system, brings inevitably to the lack of a global equilibrium
configuration that instead experience a slow evolution in time. We have therefore taken under exam the variation of the equilibrium structure while propagating in the
plasma channel due to the expansion of the front of the bunch driven by the defocusing effect of the emittance. The main feature emerging is a backward shift of the
equilibrium configuration as a function of the propagation distance. The shift appears to be linear in the initial stage, reaching then a smoother dependence after
approximately $7$ m of propagation. We found the reason of such a drift in the different behavior of the first bunch. The change in the equilibrium configuration in
fact is due to the production of a longer first bunch and therefore to a phase shift of the transverse field that it generates.\par
It is clear therefore from this analysis how in a modulated structure analogous to the one arising from the self-modulation a possible source of instability is the
divergence of the front of the bunch driven by the emittance and further efforts have to be taken in order to efficiently generate a long-lasting wakefield in such a framework.

\begin{acknowledgments}
This work was founded by DFG TR18 and EuCARD$^2$. The authors would like to thank Dr. J. P. Farmer for useful discussions on the numerical aspects developed in this
work.
\end{acknowledgments}

\providecommand{\noopsort}[1]{}\providecommand{\singleletter}[1]{#1}%

\end{document}